\documentclass[aps,prl,twocolumn,showpacs,preprintnumbers,superscriptaddress,amsmath,amssymb]{revtex4}
\usepackage{graphicx}
\usepackage{dcolumn}
\usepackage{epsfig} 

\def\Journal#1#2#3#4{{#1} {\bf #2}, #3 (#4)}

\def\NIMA{{ Nucl. Instrum. Methods} A}

\def\PRD{{ Phys. Rev.} D}

\def\YF{ Yad. Phys.} 

\newcommand{\ee}{$e^{+}e^{-}$}

\newcommand{\dob}{\ensuremath{\overline{D}{}^{0}}}

\newcommand{\RM}{\ensuremath{M_{\mathrm{recoil}}}}
\newcommand{\RMD}{\ensuremath{\Delta M_{\mathrm{recoil}}}}
\newcommand{\ISR}{\emph{ISR}}

\newcommand{\dstm}{$D^{*-}$}
\newcommand{\dstb}{$D^{(*)+}$}

\newcommand{\eetodd}{$e^{+}e^{-} \to D^{+}D^{-}$}

\newcommand{\eetoddst}{$e^{+}e^{-} \to D^{+}D^{*-}$}
\newcommand{\eetodstdst}{$e^{+}e^{-} \to D^{*+}D^{*-}$}
\newcommand{\eetodstdstb}{$e^{+}e^{-} \to D^{(*)+}D^{(*)-}$}

\newcommand{\eetoddstT}{$e^{+}e^{-} \to D^{+}D^{*-}_T$}
\newcommand{\eetoddstL}{$e^{+}e^{-} \to D^{+}D^{*-}_L$}

\newcommand{\eetodstTdstT}{$e^{+}e^{-} \to D^{*+}_TD^{*-}_T$}
\newcommand{\eetodstTdstL}{$e^{+}e^{-} \to D^{*+}_TD^{*-}_L$}
\newcommand{\eetodstLdstL}{$e^{+}e^{-} \to D^{*+}_LD^{*-}_L$}

\graphicspath{{ps}}
\begin{document}

\title{ \quad\\[0.5cm] \Large Measurement of the \eetodstdstb cross-sections}

\affiliation{Budker Institute of Nuclear Physics, Novosibirsk}
\affiliation{Chiba University, Chiba}
\affiliation{University of Frankfurt, Frankfurt}
\affiliation{Gyeongsang National University, Chinju}
\affiliation{University of Hawaii, Honolulu, Hawaii 96822}
\affiliation{High Energy Accelerator Research Organization (KEK), Tsukuba}
\affiliation{Hiroshima Institute of Technology, Hiroshima}
\affiliation{Institute of High Energy Physics, Chinese Academy of Sciences, Beijing}
\affiliation{Institute of High Energy Physics, Vienna}
\affiliation{Institute for Theoretical and Experimental Physics, Moscow}
\affiliation{J. Stefan Institute, Ljubljana}
\affiliation{Kanagawa University, Yokohama}
\affiliation{Korea University, Seoul}
\affiliation{Kyungpook National University, Taegu}
\affiliation{Swiss Federal Institute of Technology of Lausanne, EPFL, Lausanne}
\affiliation{University of Ljubljana, Ljubljana}
\affiliation{University of Maribor, Maribor}
\affiliation{Nagoya University, Nagoya}
\affiliation{Nara Women's University, Nara}
\affiliation{National United University, Miao Li}
\affiliation{Department of Physics, National Taiwan University, Taipei}
\affiliation{H. Niewodniczanski Institute of Nuclear Physics, Krakow}
\affiliation{Nihon Dental College, Niigata}
\affiliation{Niigata University, Niigata}
\affiliation{Osaka City University, Osaka}
\affiliation{Osaka University, Osaka}
\affiliation{Panjab University, Chandigarh}
\affiliation{Peking University, Beijing}
\affiliation{Princeton University, Princeton, New Jersey 08545}
\affiliation{RIKEN BNL Research Center, Upton, New York 11973}
\affiliation{University of Science and Technology of China, Hefei}
\affiliation{Seoul National University, Seoul}
\affiliation{Sungkyunkwan University, Suwon}
\affiliation{University of Sydney, Sydney NSW}
\affiliation{Tata Institute of Fundamental Research, Bombay}
\affiliation{Toho University, Funabashi}
\affiliation{Tohoku Gakuin University, Tagajo}
\affiliation{Tohoku University, Sendai}
\affiliation{Department of Physics, University of Tokyo, Tokyo}
\affiliation{Tokyo Metropolitan University, Tokyo}
\affiliation{Tokyo University of Agriculture and Technology, Tokyo}
\affiliation{Toyama National College of Maritime Technology, Toyama}
\affiliation{University of Tsukuba, Tsukuba}
\affiliation{Virginia Polytechnic Institute and State University, Blacksburg, Virginia 24061}
\affiliation{Yokkaichi University, Yokkaichi}
\affiliation{Yonsei University, Seoul}
  \author{T.~Uglov}\affiliation{Institute for Theoretical and Experimental Physics, Moscow} 
  \author{K.~Abe}\affiliation{High Energy Accelerator Research Organization (KEK), Tsukuba} 
  \author{K.~Abe}\affiliation{Tohoku Gakuin University, Tagajo} 
  \author{T.~Abe}\affiliation{High Energy Accelerator Research Organization (KEK), Tsukuba} 
  \author{H.~Aihara}\affiliation{Department of Physics, University of Tokyo, Tokyo} 
  \author{M.~Akatsu}\affiliation{Nagoya University, Nagoya} 
  \author{Y.~Asano}\affiliation{University of Tsukuba, Tsukuba} 
  \author{T.~Aso}\affiliation{Toyama National College of Maritime Technology, Toyama} 
  \author{V.~Aulchenko}\affiliation{Budker Institute of Nuclear Physics, Novosibirsk} 
  \author{A.~M.~Bakich}\affiliation{University of Sydney, Sydney NSW} 
% ...
\collaboration{The Belle Collaboration}
\noaffiliation

\begin{abstract}
We report  first measurements of \eetodstdstb\ processes far above
threshold. The cross-sections for \eetodstTdstL\ and \eetoddstT\ at
$\sqrt{s}=10.58\, \mathrm{GeV}/c^2$ are measured to be $0.55 \pm
0.03 \pm 0.05\, \mathrm{pb}$ and $0.62 \pm 0.03 \pm 0.06\,
\mathrm{pb}$, respectively. We set upper limits on the cross-sections
for \eetodstTdstT, \eetodstLdstL, \eetoddstL\ and \eetodd\ processes.
The analysis is based on $88.9\,\mathrm{fb}^{-1}$ of data collected by
the Belle experiment at the KEKB $e^{+}e^{-}$ asymmetric collider.

\end{abstract}
\maketitle

\tighten
{\renewcommand{\thefootnote}{\fnsymbol{footnote}}}
\setcounter{footnote}{0}

\noindent
The processes $e^{+}e^{-} \to D^{(*)} \overline{D}{}^{(*)}$, with no
extra fragmentation particles in the final state, have not previously
been measuered for energies $\sqrt{s} \gg 2M_{D}$.  The cross-sections
for these processes can be computed once the charmed meson form
factors are determined for the appropriate value of momentum transfer,
$q^2 \equiv s$.  In the HQET approach based on the heavy-quark spin
symmetry, the heavy meson form factors can be expressed in terms of a
universal form factor, called the Isgur-Wise function.  However, for
large $q^2$, the leading-twist contribution, which violates
heavy-quark spin symmetry, becomes dominant~\cite{neubert}.
For an intermediate range of momentum
transfer, the Isgur-Wise contribution (having subleading twist) is
also important. A calculation that takes these effects into
account \cite{neubert} predicts that the cross sections for $e^+e^-\to
D\overline{D}{}^*$ and $e^+e^-\to D^*_L\overline{D}{}^*_T$ are
equal to each other, and are about $2.5$ pb at
$\sqrt{s}\sim10.6\,\mathrm{GeV}$ (the subscripts indicate
longitudinal [L] and transverse [T] polarizations of the $D^*$).
The cross-section for $e^+e^-\to D\bar{D}$ is expected to be
suppressed by a factor of $\sim10^{-3}$. The cross-section
predictions are subject to large uncertainties ($\sim$ a factor of
3), because the theoretical formulae contain many poorly known
parameters; the prediction that the $D\overline{D}{}^*$ and
$D^*_L\overline{D}{}^*_T$ cross-sections should be equal is
expected to be robust \cite{private}. Recently, new calculations in the
framework of the constituent quark model have become available \cite{kitaicy}.


In this paper, we present the first measurement of the \eetodstdst\
and \eetoddst\ cross-sections and polarizations at $\sqrt{s} \sim
10.6\, \mathrm{GeV}$.  We also set an upper limit on the cross-section
for \eetodd.  The present study is limited to final states that
contain charged $D^{(*)}$ mesons only. Since the contribution of the
electromagnetic current coupled to light quarks is negligible compared
to that for heavy quarks, the neutral and charged charm meson
cross-sections are expected to be the same~\cite{neubert}.

The analysis is based on $88.9\,\mathrm{ fb}^{-1}$ of data at the
$\Upsilon(4S)$ resonance and nearby continuum, collected with the
Belle detector~\cite{Belle} at the KEKB asymmetric energy \ee 
collider~\cite{KEKB}.  We select well-reconstructed tracks consistent with
originating from the interaction region as charged pion
candidates. Those passing particle identification cuts based on
$dE/dx$, aerogel \v{C}erenkov, and time-of-flight
information~\cite{Belle} are selected as charged kaon candidates.  We
then reconstruct $D^0$ and $D^+$ mesons in the decay modes $D^0 \to
K^- \pi^+$, $D^0 \to K^- \pi^+ \pi^+ \pi^-$ and $D^+ \to K^- \pi^+
\pi^+$. The selected combinations are constrained to a common vertex,
and quality cuts are imposed on the vertex fit to reduce the
combinatorial background. A $15\,\mathrm{MeV}/c^2$ interval around the
nominal $D$ masses is used to select $D^0 \to K^- \pi^+$ and $D^+ \to
K^- \pi^+ \pi^+$ candidates; for the $D^0 \to K^- \pi^+ \pi^+ \pi^-$
decay mode the signal window is chosen to be $ 10\,\mathrm{MeV}/c^2$
around the nominal $D^0$ mass ($\sim2\,\sigma$ each case). The
selected $D$ candidates are then subjected to a mass and vertex
constrained fit to improve their momentum resolution.  The $D^{*+}$
mesons are reconstructed in the $D^0 \pi^+$ decay mode. The mass of
the $D^0 \pi^+$ combination is required to be within a $
2\,\mathrm{MeV}/c^2$ ($\sim 3 \,\sigma$) mass interval around the
nominal $D^{*+}$ mass.

The processes $e^+ e^- \to D^{(*)+} D^{(*)-}$ can be identified by
energy-momentum balance in fully reconstructed events that contain
only a pair of charm mesons.  However, the small charm meson
reconstruction efficiency of the studied channels results in a tiny
total efficiency in this case. Because of the simple two-body
kinematics, it is sufficient to reconstruct only one of the two
charmed mesons in the event to identify the processes of interest. For
simplicity, we refer to the fully reconstructed $D$ meson as the
\dstb, and the other as the \dstm; the charge-conjugate modes are
included in the analysis. We choose the mass of the system recoiling
against the reconstructed \dstb\ ($\RM(D^{(*)+})$) as a discriminating
variable: $
\RM(D^{(*)+})=\sqrt{(\sqrt{s}-E_{D^{(*)+}})^2-\vec{p}_{D^{(*)+}}^{~2}},
$ where $\sqrt{s}$ is the total center of mass (CM) energy, and
$E_{D^{(*)+}}$ and $\vec{p}_{D^{(*)+}}$ are the CM energy and momentum
of the reconstructed $D^{(*)+}$. For the signal, a peak in the \RM\
distribution around the nominal  $D^-$ or $D^{*-}$ mass is
expected. This method provides a significantly increased efficiency,
but also a higher background, in comparison with full event
reconstruction. For the $e^+ e^- \to D^+ D^{*-}$ and $e^+ e^- \to
D^{*+} D^{*-}$ processes we find a better compromise between higher
statistics and smaller background: the first $D^{(*)+}$ is fully
reconstructed, while the recoiling $D^{*-}$ is required to decay into
$\dob \pi^-_{slow}$. The reconstructed $\pi^-_{slow}$ provides
extra information that allows us to reduce the background to a
negligible level using the difference between the masses of the
systems recoiling against the $D^{(*)+} \pi^-_{slow}$ combination, and
against the $D^{(*)+}$ alone, $ \RMD\equiv \RM(D^{(*)+})-\RM(D^{(*)+}
\pi^-_{slow})$. The variable \RMD\ peaks around the nominal
$D^{*-}\!\!-\!\!\dob$ mass difference with a resolution of
$\sigma_{\RMD}\!\sim\!  1\,\mathrm{MeV}/c^2$ as found by Monte Carlo
(MC) simulation.  For \eetodstdst\ and \eetoddst\ we require \RMD\ to
be within a $\pm 2\,\mathrm{MeV}/c^2$ interval around the nominal
$M_{D^{*-}} -M_{\dob}$ mass difference.

\begin{figure}[htb]
\begin{center}
\hspace*{-0.1cm}\epsfig{file=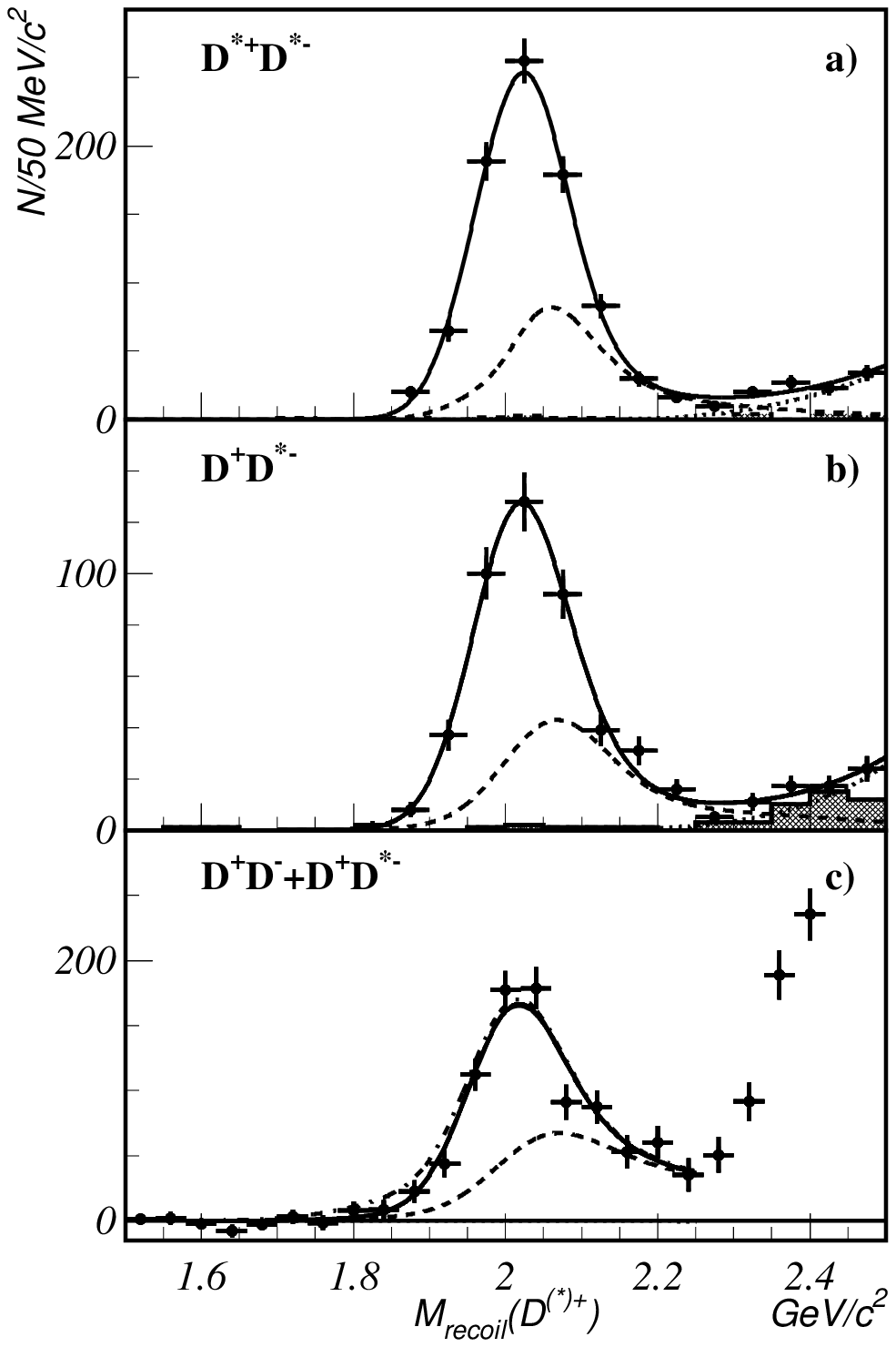,width=0.38\textwidth}
\end{center}
\caption{ a) $\RM(D^{*+})$ and b) $\RM(D^+)$ after applying the \RMD\
 requirement. Points show the \RMD\ signal region; hatched histograms
 show the \RMD\ sideband; solid lines represent the fits described in
 the text; dashed lines show the contribution due to events with \ISR\
 photons of significant energy; dotted lines are the expected
 background contribution. c) $\RM(D^+)$ after $D^+$ sideband
 subtraction without requiring an extra $\pi^-_{slow}$ in the event.
}
\label{dstdst}
\end{figure}

The $\RM(D^{*+})$ and $\RM(D^+)$ distributions are shown in
Figs.~\ref{dstdst}a and~\ref{dstdst}b, respectively. Clear signals are
seen around the nominal $D^{*-}$ mass in both cases. The higher recoil
mass tails in the signal distributions are due to initial state
radiation (\ISR). The \RM\ distributions for events in the \RMD\
sideband ($0.150 \, \mathrm{MeV}/c^2 < \RMD < 0.154 \,
\mathrm{MeV}/c^2$) are shown as the hatched histogram (barely
visible in Fig.~\ref{dstdst}a due to its small size).

The backgrounds in the region $\RM < 2.1 \, \mathrm{GeV}/c^2$ are
negligible for both processes, so we consider this interval as the
signal region. There are three possible background sources:
\begin{itemize} 
\item[I~] incorrectly reconstructed $D^{*+}$ or $D^+$;
\item[II~] $e^+ e^- \to D^{(*)+} D n \pi^-$ ($n \ge 0$), where the
$\pi^-_{slow}$ is not produced from $D^{*-}$ decay (and can be either
from fragmentation or from unreconstructed $D$ decay), and thus
produces no peak in the \RMD\ distribution;
\item[III~] $e^+ e^- \to D^{(*)+} D^{*-} n \pi $, where $n \ge 1$.
\end{itemize}
First we consider the process \eetodstdst. To estimate background (I)
numerically, we count the entries in the signal region for $D^0 \pi^+$
combinations taken from the $D^{*+}$ mass sideband ($2.016\,
\mathrm{GeV}/c^2 < M_{D^0\pi^+} <2.020\, \mathrm{GeV}/c^2$). Three
events are found in the data, while the MC predicts a contribution of
2.5 events from the signal process due to non-Gaussian tails in the
$M_{D^{*+}}$ resolution function. The signal MC is normalized to the
number of entries in the $\RM(D^{*+}) < 2.1 \, \mathrm{GeV}/c^2$
region in the data.  Background (II) is estimated using the \RMD\
sideband ($0.150 \, \mathrm{MeV}/c^2 < \RMD < 0.154 \, \mathrm{MeV}/c^2$). In the signal
region 8 events are found; 4 events are expected according to MC from
the signal process because of significant energy \ISR. Thus
backgrounds (I) and (II) are estimated to be smaller than $5$ and $10$
events at the $90\%~\mathrm{CL}$, respectively.  The remaining
background (III) can result in peaks in both the $M(D^{*+})$ and \RMD\
distributions, but has a threshold in the \RM\ distribution at $M_{D^{*+}}
+ M_{\pi^0} = 2.15 \, \mathrm{GeV} / c^2$, which is
$\!\sim\!1\,\sigma$ higher than the chosen \RM\ signal interval. To
estimate the residual background (III) contribution in the signal
region we perform a fit to the \RM$(D^{*+})$ distribution. The signal
function is determined from the MC simulation and parametrized as the
sum of a core Gaussian and an asymmetric function representing the
case when the studied process is accompanied by radiative photon(s)
with significant energy ($E_{ISR}>10\,\mathrm{MeV}$).  The
\RM$(D^{*+})$ resolution due to detector smearing and the signal
function offset are left as free parameters in the fit to check the
agreement with the MC predictions. The background (III) distribution
is parameterized by a threshold function, $ \alpha
(\RM(D^{*+})-M(D^{*-})_{PDG} -M(\pi^0)_{PDG})^{\beta}$, convolved with
the detector resolution, where $\alpha$ and $\beta$ are free
parameters. The fit results are shown in the Fig.~\ref{dstdst}a as a
solid curve; the dashed line shows the contribution of the studied
process with significant energy \ISR\ photons
($E_{ISR}>10\,\mathrm{MeV}$). The dotted line represents the expected
background (III) distribution. The signal yield is found to be $815\pm
28$ events in the whole fit region ($676\pm24$ events in the signal region $\RM < 2.1 \, \mathrm{GeV}/c^2$). The \RM\ resolution $\sigma=56.1
\pm 2.2 \, \mathrm{MeV}/c^2$ is found to be in excellent agreement
with the MC expectation ($56.4\, \mathrm{MeV}/c^2$), and the shift of
the signal peak position in the data with respect to the MC position
is found to be consistent with zero ($0.6 \pm 2.5\,\mathrm{MeV}/c^2$).
The contribution from background (III) in the signal region is
estimated from this fit to be less than $2$ events at the $90\%$ CL.

For the process $e^+ e^- \to D^+ D^{\ast -}$, proceeding in a
similar way, we find a signal yield of $423\pm20$ events in the whole fit region ($360\pm17$ events in the signal region $\RM < 2.1 \, \mathrm{GeV}/c^2$), with
backgrounds (I), (II) and (III) smaller than $8$, $6$ and $2$ events at
the $90\%$ CL, respectively. 
Finally we estimate the total background in
the $\RM < 2.1\, \mathrm{GeV}/c^2$ interval to be smaller than $13$
and $10$ events for the \eetodstdst\ and \eetoddst\ processes,
respectively, which is of the order of $1\%$ of the signal. We therefore
assume that all events in the interval $\RM < 2.1\, \mathrm{GeV}/c^2$ are
signal, and include the possible background contribution in the
systematic error.

\begin{figure}[htb]
\begin{center}
\epsfig{file=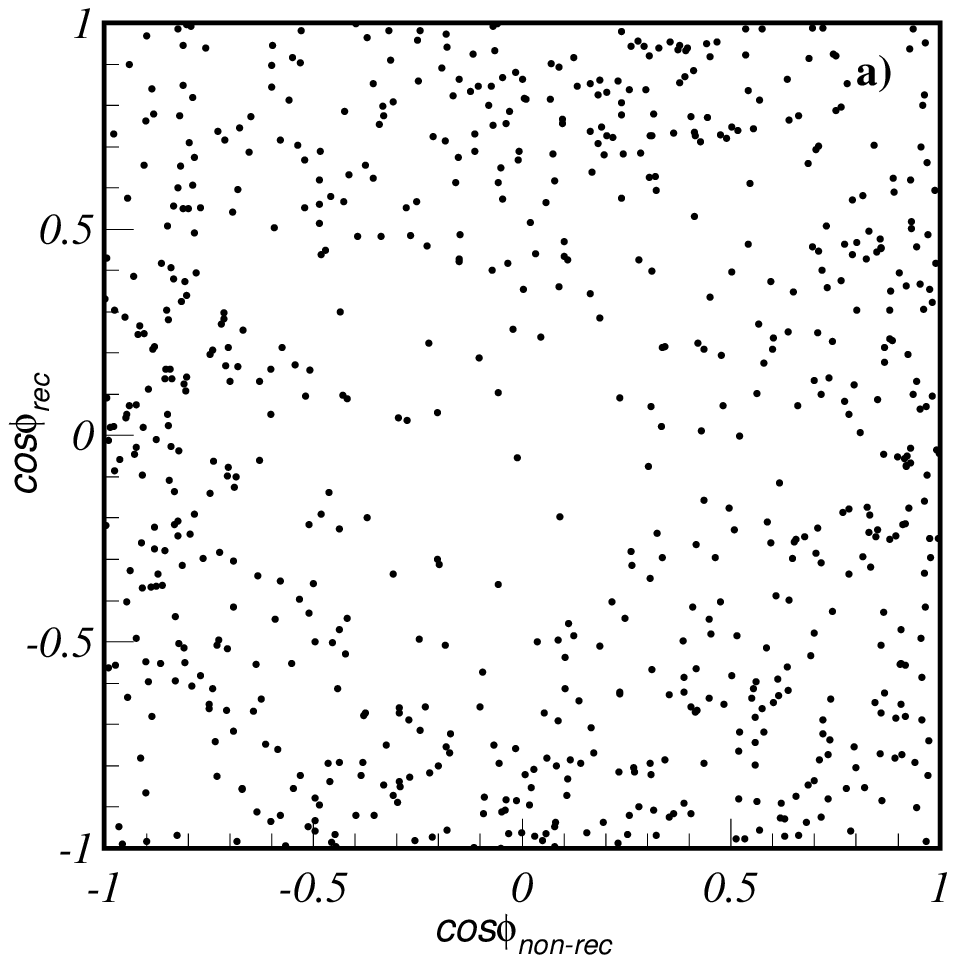, width=0.38\textwidth}
\epsfig{file=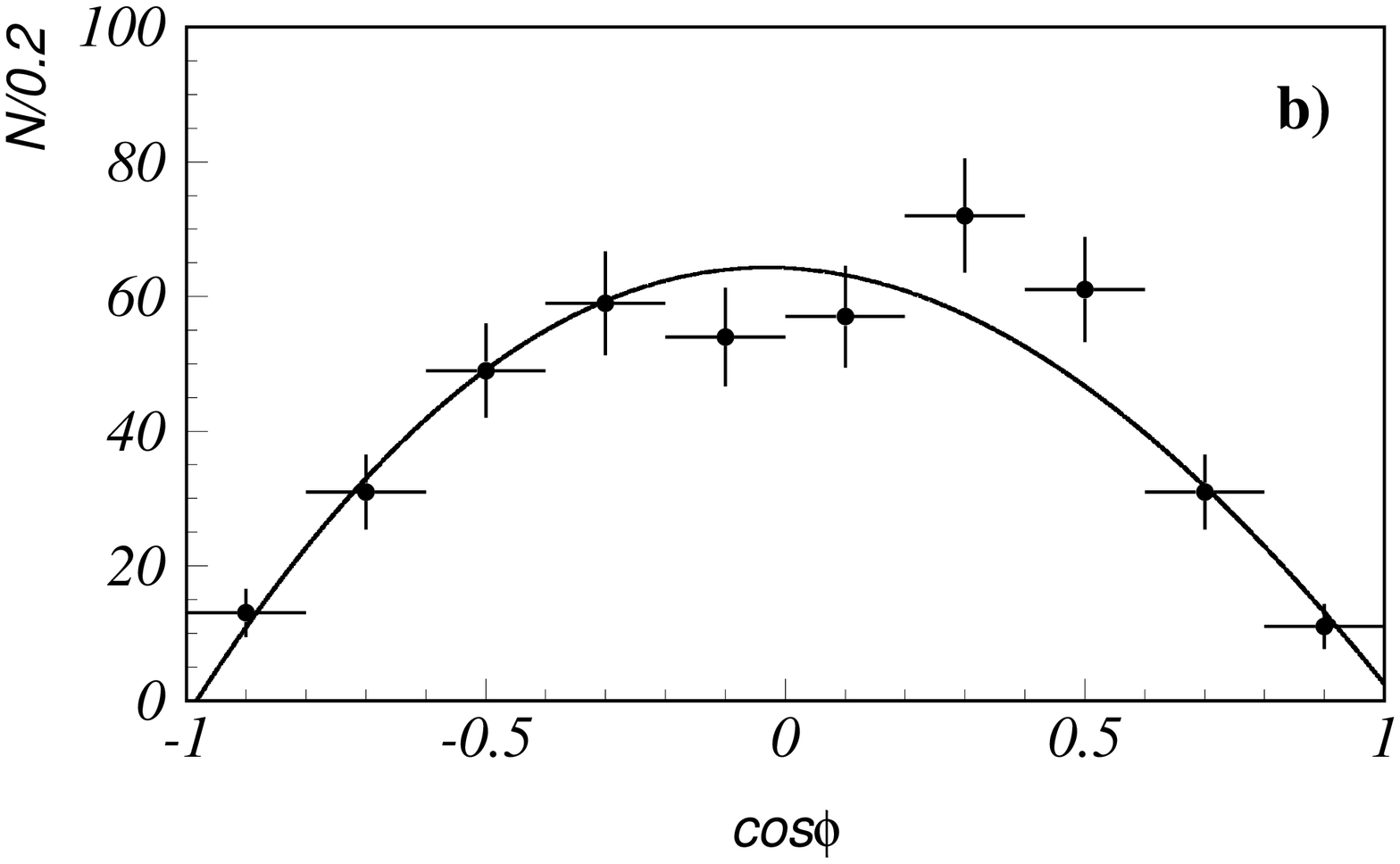, width=0.38\textwidth, height=0.25\textwidth}
\end{center}
\caption{a) A scatter plot of $\cos(\phi_{D^*_{rec}})$ {\emph {vs.}}
$\cos(\phi_{D^*_{non-rec}})$ for \eetodstdst events. b) $D^{*-}$ meson helicity
angle distribution for \eetoddst\ signal candidates. The curve
represents the fit described in the text.}
\label{heli}
\end{figure}

Since the reconstruction efficiency depends on the production and
$D^{*\pm}$ helicity angle distributions, we perform an angular
analysis before computing cross-sections. The helicity angle of the
non-reconstructed $D^{*-}$ is calculated assuming two-body
kinematics. A scatter plot of the helicity angles for the two
$D^{*}$-mesons from \eetodstdst\ ($\cos\phi(D^*_{rec})$ \emph {vs.}
$\cos\phi(D^*_{non-rec})$) for the signal region is shown in
Fig.~\ref{heli}a. This two dimensional distribution is fitted by a sum
of three functions corresponding to the $D^*_T D^*_T$, $D^*_T D^*_L$
and $D^*_L D^*_L$ final states, obtained from the MC simulation. The
fit finds $6^{+15}_{-13}$, $708\pm36$ and $4^{+18}_{-17}$ events
associated with $D^*_T D^*_T$, $D^*_T D^*_L$ and $D^*_L D^*_L$ final
states, respectively.  Figure~\ref{heli}b shows the $D^{*-}$ meson
helicity distribution for \eetoddst. A fit finds $433\pm24$ and
$-1.5\pm2$ events corresponding to $DD^*_T$ and $DD^*_L$,
respectively.
We conclude that in \eetodstdst\ and \eetoddst\ production the final
states are saturated by $D^*_T D^*_L$ and $DD^*_T$. ($D D_T^*$ is
required by angular momentum and parity conservation for $e^+ e^- \to
D^+ D^{*-}$ via a virtual photon; the $D_T^* D_L^*$ result is
non-trivial.) The production angle distributions for \eetodstTdstL\
and \eetoddstT\ are, therefore, fixed to be $1+\cos^2{\theta}$ in both
cases. As a cross-check we study the production angle distributions
for $D^{*+}$ from \eetodstdst\ and $D^+$ from \eetoddst\
processes. After correction for reconstruction efficiency in bins of
the production angle we fit the production angle distributions with
the function $(1 + A \cos^{2}{\theta})$.  The parameters $A$ are found
to be equal to $0.79^{+0.34}_{-0.30}$ and $2.3_{-0.7}^{+0.8}$ for the
two processes, which are in agreement with the expected value $A=1$
for both processes.

To calculate the Born cross-section for the studied processes we
determine the fraction of events in the signal region with
an \ISR\ photon energy smaller than the chosen cutoff
($E_{\text{cutoff}}=10\,\mathrm{MeV}$) using a MC simulation.
 In the MC we assume a $1/q^6$
dependence of the ratio of the cross section to
$\sigma(e^+e^-\to\mu^+\mu^-)$ as predicted by Ref.~\cite{neubert}. We also
try a $1/q^2$ dependence (corresponding to flat form-factors) and
include the resulting shift 
%
%
of the event fraction ($<1.3\%$) in the systematic
error. The reconstruction efficiencies are determined from MC
simulation. The Born cross-sections for $e^+ e^- \to D^{(*)+}D^{*-}$
are calculated according to the formulae in Ref.~\cite{fadin} and are
listed in Table~\ref{summary}. The final result is
independent of the choice of $E_{\text{cutoff}}$.  The sources of systematic
error are summarized in Table~\ref{systematics}. The dominant
contributions are from the uncertainties in tracking efficiency and
$D^{(*)}$ branching ratios.

\begin{table}[h!]
\begin{center}
\begin{tabular}{|l|c|}
\hline 
\ Process  & $\sigma_{Born}$ (pb) \\
 \hline
\ \eetodstTdstT \ \ & \ \ $<0.02\,@\,90\% $ CL \ \ \\
\ \eetodstTdstL &  $0.55\pm 0.03\pm 0.05$ \\
\ \eetodstLdstL &  $< 0.02\, @\,90\% $ CL  \\
\hline
\ \eetoddstL &  $< 0.006\,@\,90\%$ CL \\
\ \eetoddstT &  $0.62\pm0.03\pm 0.06$ \\ 
\hline
\ \eetodd &  $<0.04\,@\,90\%$ CL \\
\hline
\end{tabular}
\end{center}
\caption{The Born cross-section results. The first error is statistical and the second one is systematical.}
\label{summary}
\end{table}

\begin{table}[htb]
\caption{Summary of systematic errors for the \eetodstdst\ and
\eetoddst\ measurements.}
\begin{center}
\begin{tabular}{|l|c|c|}
\hline
\ Source  &\ \eetodstdst \ &\ \eetoddst \ \\
\hline
\ Tracking efficiency \   & 7\%   & 5\%  \\
\ Identification          & 2\%   & 2\%  \\
\ Backgrounds             & $~^{+0}_{-1.6}\%$   &  $~^{+0}_{-2.5}\%$ \\
\ Form-factor shape       & 1.3\%   & 1.3\%  \\
\ Luminosity              & 1\%   & 1\%  \\
\ $\mathcal{B}(D^{(*)})$  & 4\%   & 8\%  \\
\hline
\ Total                   & 9\%   & 10\% \\
\hline
\end{tabular}
\end{center}
\label{systematics}
\end{table}

We search for the process \eetodd\ by studying the mass spectrum of
the system recoiling against the reconstructed $D^+$ without requiring an extra
soft pion in the event. In the \eetodstdst\ and \eetoddst\ analyses,
backgrounds are strongly suppressed by the  \RMD\ cut, which is
not applicable for the \eetodd\ search; without this requirement the
combinatorial non-$D^+$ background is significant. We use $D^+$ mass
sidebands ($20 \, \mathrm{MeV}/c^2< \left| M_{K\pi\pi}-M_D \right| <35 \,
\mathrm{MeV}/c^2$) to extract the \RM\ distribution for the
combinatorial background. Figure~1c shows the \RM$(D^+)$ distribution 
after $D^+$ mass sideband subtraction. We fit this distribution with the sum of
two signal functions corresponding to $D^-$ and $D^{*-}$ peaks and a
background function.  The latter is a threshold function, $\alpha
(x-M(D^{-})_{PDG} -M(\pi^0)_{PDG})^{\beta}$, convolved with the
detector resolution, where $\alpha$ and $\beta$ are free
parameters. For the fit we use only the region $\RM <
2.25\,\mathrm{GeV}/c^2$, because of a possible contribution of $e^+
e^- \to D^{(*)} D^{**}$ at higher \RM.  The fit finds $13\pm 24$
events in the $D^-$ peak and $935\pm 42$ in the $D^{*-}$ peak.  The
fit function is shown in Fig.~1c as the solid line; the dashed line
shows the contribution of events with \ISR\ photons of significant
energy (larger in this case due to the absence of the \RMD\ cut); and
the dotted line represents the case where the contribution of \eetodd\
is set at the value corresponding to the $90\%$ CL upper limit.  The
reconstruction efficiencies for \eetodd\ and \eetoddst\ are found from
MC. The production angle distribution for \eetodd\ is assumed to be
proportional to $\sin^2{\theta}$, while the production angle for
\eetoddst\ is fixed from the study with the \RMD\ requirement. 
 The \eetoddst\
Born cross-section is calculated to be $0.54 \pm0.04\,
\mathrm{pb}$. In this method the systematic uncertainty in the signal
yield is larger than in the \RMD\ method due to significant $e^+ e^-
\to D^+ D \pi$ background under the peak, which can only be
extrapolated from the higher \RM\ region with large uncertainties. For
the \eetodd\ Born cross-section we set an upper limit of $0.04\,$pb at
the $90\%$ CL.

The relative sizes of the measured cross-sections agree with the
predictions of Ref.~\cite{neubert}: $\sigma(e^+e^-\to D^{*+}_T D^{*-}_L)$ and
$\sigma(e^+e^-\to D^+D^{*-}_T)$ are similar, while
$\sigma(e^+e^-\to D^+D^-)$ is much smaller. $e^+e^-\to
D^{*+}D^{*-}$ production is saturated by the $D^{*+}_T D^{*-}_L$
final state, also as expected.  The absolute cross-sections are
smaller than those of \cite{neubert} by a factor of 4, which is comparable to
the theoretical uncertainty. Recent calculations based on the
constituent quark model \cite{kitaicy} reproduce the $D^{*+}D^{*-}$ and $D^+
D^{*-}$ cross-sections very well, but predict $\sigma(e^+e^-\to
D^+D^-) = 0.1\,\mathrm{pb}$, somewhat larger than our limit. The
predicted $D^{*+}_L D^{*-}_T$ fraction in $e^+e^-\to D^{*+}D^{*-}$
production, $65\%$ \cite{kitaicy}, is smaller than we observe.
%
%

In summary, we report the first measurement of the cross-sections for
the \eetodstTdstL\ and \eetoddstT\ processes at $\sqrt{s}=10.6\,
\mathrm{GeV}$ and set upper limits on the \eetodstTdstT,
\eetodstLdstL, \eetoddstL\ and \eetodd\ cross-sections.

We are grateful to A.G.Grozin for useful discussion and  comments on
theoretical issues.

We wish to thank the KEKB accelerator group for the excellent
operation of the KEKB accelerator.
We acknowledge support from the Ministry of Education,
Culture, Sports, Science, and Technology of Japan
and the Japan Society for the Promotion of Science;
the Australian Research Council
and the Australian Department of Education, Science and Training;
the National Science Foundation of China under contract No.~10175071;
the Department of Science and Technology of India;
the BK21 program of the Ministry of Education of Korea
and the CHEP SRC program of the Korea Science and Engineering Foundation;
the Polish State Committee for Scientific Research
under contract No.~2P03B 01324;
the Ministry of Science and Technology of the Russian Federation;
the Ministry of Education, Science and Sport of the Republic of Slovenia;
the National Science Council and the Ministry of Education of Taiwan;
and the U.S.\ Department of Energy.


\begin{thebibliography}{99}

\bibitem{neubert} A.G.~Grozin, M.~Neubert, \Journal{\PRD}{55}{272}{1997}.

\bibitem{private} A.G.~Grozin, private communications.

\bibitem{kitaicy} K.-Y.~Liu {\it et al.}, hep-ph/0311364.

\bibitem{Belle} A.~Abashian {\it et al.}, \Journal{\NIMA}{479}{117}{2002}.

\bibitem{KEKB} S.~Kurokawa and S.~Kikutani \Journal{\NIMA}{499}{1}{2003}.

\bibitem{fadin} E.A.~Kuraev, V.S.~Fadin,
\Journal{\YF}{41}{733-742}{1985}.

\end{thebibliography}
\end{document}